\documentclass[
reprint,
superscriptaddress,
amsmath,amssymb,
prstab,
floatfix
]{revtex4-2}

\usepackage{placeins}
\usepackage{enumerate}
\usepackage{csquotes}

\usepackage{graphicx}

\usepackage[dvipsnames]{xcolor}
\usepackage{amsmath}

\usepackage{siunitx}
\DeclareSIUnit\bar{bar}
\DeclareSIUnit\var{VAR}

\usepackage{tablefootnote}

\graphicspath{ {./images/} }

\begin{document}

\preprint{AIP/123-QED}

\title{High Power Fast Frequency Modulation }

\author{Ilan \surname{Ben-Zvi}}
\email{Ilan.Ben-Zvi@StonyBrook.edu}
\affiliation{Physics and Astronomy Department, Stony Brook University, NY USA}

\author{Nicholas \surname{Shipman}}
\affiliation{Helmholtz Centre for Materials and Energy, Berlin, Germany}
\email{nicholas.shipman@helmholtz-berlin.de}

\date{\today}

\begin{abstract}
A fast and highly efficient frequency modulation at a high power level is described. The system incorporates ferroelectric phase shifters and a magic-T or a circulator. A magnetron may be considered as a potential application. The magnetron output may be converted to a selected reference frequency with negligible insertion loss. The method also allows simultaneous amplitude and phase control. 
\end{abstract}

\maketitle


\section{Introduction}\label{sec:intro}
Some microwave power applications could benefit from an efficient scheme to change the frequency of a possibly unstable high-power source. For example, magnetrons are attractive RF sources due to their high wall-plug efficiency and low cost. However, their adoption in accelerator physics has lagged due to their poor frequency stability, prompting numerous approaches to solve this issue over the past few decades \cite{Injection},\cite{vyas2016review}. This paper proposes a novel technique to shift the frequency of a high-power microwave source in a feed-forward scheme, as well as allowing phase and amplitude feedback.

The key idea is to add a saw tooth phase variation to the source signal before it is sent to the load, this requires extremely fast, low loss, high power phase shifters.  Recent developments with Ferro-Electric Fast Reactive Tuners (FE-FRTs) have now made this technique viable\cite{ben2024conceptual}.

Good progress towards stabilizing magnetron frequencies has already been made with injection locking techniques\cite{kazakevich2018resonant}, \cite{WangSRF23}.  However, in such schemes, the injection power needed increases with the required locking bandwidth and can be significant, impacting the overall system efficiency.  Depending on the application and required frequency shift, the techniques proposed in this paper could be used instead of an injection locking scheme, leading to gains in system efficiency and stability.

Many different RF circuits capable of performing the proposed technique can be envisioned.  Three such circuits are presented in secs.~\ref{sec:magic},\ref{sec:Circulator} and\ref{sec:transmission}.  As FE-FRTs have very low insertion losses, and can handle high continuous power, all three of the presented methods would allow high efficiency frequency conversion of high power signals.  The first is a somewhat complicated circuit, but based on a well established phase and amplitude control technique \cite{Valuch2004}, \cite{yakovlev20061}.  The following two are somewhat simpler.  The increased simplicity is owed to an alternative method of amplitude control, described in sec.~\ref{sec:switch timing}, which is practicable due to the rapid phase control enabled by FE-FRTs.

\section{Frequency Modulation}\label{sec:freq}

The objective is to efficiently convert the microwave power of the source at frequency $\omega$ to power at the frequency required by the load $\omega _0$, which differ by some value $\Omega$ where:
\begin{equation}\label{eq:Omega}
    \Omega\equiv\omega-\omega_0
\end{equation}

$\Omega$ is small relative to $\omega$ and may vary in time due to a change in $\omega$.  A reference master oscillator provides a low-level RF signal at $\omega _0$.  The output must be phase locked to the reference and amplitude feedback may also be desirable.

The source signal $S_M$ can thus be written:
\begin{equation}
    S_M =V_M e^{j \omega t}=V_M e^{j \omega _0 t} e^{j \Omega t}
      \label{eq:1}
\end{equation}
with $V_m$ a real positive amplitude.

A circuit containing FE-FRT(s)\cite{ben2024conceptual} is designed such that the signal sent to the load, $S_L$, is:
\begin{equation}\label{eq:SL_given}
    S_L = \Gamma S_M = \Gamma V_M e^{j \omega _0 t} e^{j \Omega t}
\end{equation}

Where $\Gamma$ is the reflection coefficient of the FE-FRT and, as the losses are small, has a magnitude close to unity such that it can be written as:
\begin{equation}
\Gamma = |\Gamma|e^{j\theta}\approx e^{j\theta}
    \label{eq:2}
\end{equation}

In the special case where no amplitude modulation is required the desired load signal is:
\begin{equation}\label{eq:SL_required}
    S_L = V_Me^{j\omega_0t}
\end{equation}

Comparing eq.~\ref{eq:SL_given} to eq.~\ref{eq:SL_required} reveals the requirement:
\begin{equation}\label{eq:Gamma_is_eOmegat}
    \Gamma = e^{-j\Omega t}
\end{equation}

Therefore from eq.~\ref{eq:2}, the phase shift introduced by the FE-FRT must be:
\begin{equation}\label{eq:theta}
    \theta = -\Omega t
\end{equation}

\begin{figure}[tb]
    \centering
\includegraphics[width=0.9\columnwidth]{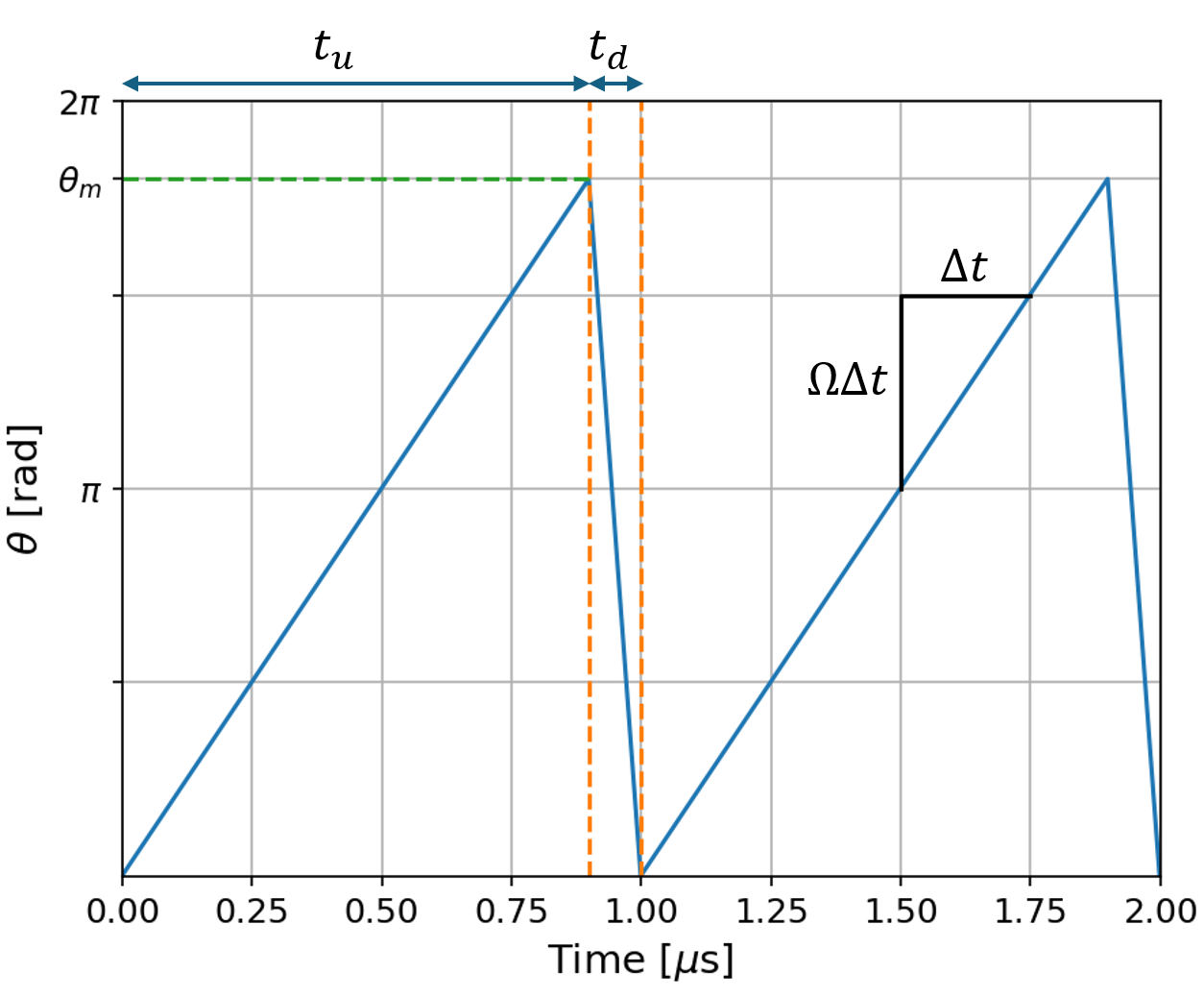}   
    \caption{Example of the phase waveform (blue) for a case where $\Omega = 1\,\textrm{MHz}$.  $t_u$ and $t_d$ are marked by the orange dashed lines.  $\theta_m$, the maximum phase shift provided by the FE-FRT, is marked by the red dashed line.  $t_d$ has been exaggerated for clarity.}
     \label{fig:sawtooth}
\end{figure}

In practice, the FE-FRT can only provide a finite range of phase shift and must therefore be periodically reset.  The  waveform of the phase $\theta$ is thus a sawtooth as shown in fig.~\ref{fig:sawtooth}.  The ramp of the sawtooth has a gradient $\Omega$ and occurs over a time $t_u$ and the reset happens over a much shorter time $t_d$.  The period $\tau$ is thus:
\begin{equation}\label{eq:tau}
\tau = t_u + t_d
\end{equation}
the rep-rate $f_{rep}$ is:

\begin{equation}
f_{rep} =\frac{1}{\tau}
\label{eq:4}
\end{equation}
and the maximum phase shift provided by the FE-FRT, $\theta_m$ is:
\begin{equation}\label{eq:theta_m}
\theta_m = -\Omega t_u
\end{equation}

During the brief period $t_d$, when the derivative of $\theta$ is not $\Omega$, the source is converted to a much different frequency than $\omega_0$. Therefore, the fraction $t_d / \tau \ll 1$ of the source power is wasted.  Ignoring for now other loss mechanisms this implies the maximum efficiency of the power conversion $\eta$ will be:
\begin{equation}
\eta \leq 1 - \frac{t_d}{\tau}
\end{equation}
Clearly therefore, it is desirable to keep $t_d$ as small as possible.

Many applications require the phase sent to the load to be locked to the phase of the reference; this imposes extra conditions on $\tau$ and $\theta_m$ which are now examined.

Define the difference in phase between the reference signal and signal sent to the load as $\Delta\Phi(t)$.  The phase shift $\theta(t)$ is added through the FE-FRT to the source signal before it is sent to the load. Without loss of generality, define for $t=0$:
\begin{equation}\label{eq:Phi_0_is_0}
\Delta\Phi(0) = 0
\end{equation}
and:
\begin{equation}
\theta(0) = \theta(\tau)=0
\end{equation}

During the ramp-up and the reset, the phases of the source and the reference advance at their respective rates $\omega t$ and $\omega_0 t$.  Therefore, after the reset (where $\theta(\tau) = 0$):
\begin{equation}
\Delta\Phi(\tau)=\omega \tau -\omega_0 \tau=\Omega \tau 
\label{eq:7}
\end{equation}

Given eq.~\ref{eq:Phi_0_is_0}, phase locking of the signal sent to the load to the reference signal implies:
\begin{equation}
\Delta\Phi(\tau)= 0\,\,(\textrm{mod}\,2\pi)
\end{equation}

Therefore, from eq.~\ref{eq:7}, $\tau$ must be set to:
\begin{equation}\label{eq:tau_Omega}
\tau = \frac{2\pi n}{\Omega}
\end{equation}
or equivalently:
\begin{equation}
f_{rep} = \frac{\Omega}{2\pi n}
\end{equation}
where n is an integer.  From eq.~\ref{eq:tau}, eq.~\ref{eq:theta_m} and eq.~\ref{eq:tau_Omega} it can be shown that the FE-FRT must be able to provide a maximum phase shift $\theta_m$ of:
\begin{equation}\label{eq:theta_m_tau}
\theta_m = 2\pi n\left(\frac{\tau-t_d}{\tau}\right)
\end{equation}

It is difficult to design a high performant FE-FRT capable of providing a phase shift greater than $2\pi$ and hence, in practice, it would be likely that $n=1$.

To achieve the required phase waveform a Digital Signal Processor (DSP) is used to measure the difference between $\omega_0$ and $\omega$ and generate appropriate signals to control the voltage applied to FE-FRTs \cite{ben2024conceptual}, shown as phase shifters in Figures \ref{magic},  \ref{circ} and \ref{trans_mode}. 

\section{Phase and Amplitude Feedback}\label{sec:feedback} 
In addition to the frequency conversion, the scheme allows for phase feedback and amplitude feedback, to provide an exceptional quality drive to the load.  The outgoing signal of the frequency converter is compared to the desired amplitude and phase by the control system which generates error signals and calculates appropriate corrections.

Two types of feedback are possible.  ``Direct feedback" applies a correction voltage directly to the phase shifter on top of the frequency modulation voltage.  Whereas ``Feedback by switch timing" works by allowing the time at which a reset occurs to be varied,which allows corrections to be made at a maximum frequency of approximately $f_{rep}$. 

\subsection{Direct feedback}\label{sec:direct}

 Direct feedback allows for continuous feedback and adjustment of the phase shift from the FE-FRT(s) on timescales smaller than $\tau$ at the cost of some extra power required from the driving circuitry.  In all the circuits presented in the following sections, direct feedback would allow extremely rapid control of the phase of the signal sent to the load.  Crucially, this continuous and rapid phase control would allow corrections of any non-linearity in the FE-FRT phase shift as a function of applied voltage.

Direct feedback, when applied in the magic-T scheme (sec.~\ref{sec:magic}), would also allow for regulation of the amplitude in addition to the phase via controlling the difference and sum of the phase shift in the two FE-FRTs respectively.

Direct feedback could be achieved using a voltage controlled current source as described in sec.~\ref{sec:driver}, which would allow for small modulations of the DC current used to achieve the saw tooth waveform required for uncorrected frequency modulation.

\subsection{Feedback by switch timing}\label{sec:switch timing}

Feedback through switch timing can be implemented by modifying the reset point of the sawtooth waveform shown in Fig.~\ref{fig:sawtooth}. Equivalently, this corresponds to adjusting $t_u$, the time during which the phase shift is ramping.  Decreasing $t_u$ will shift the waveform to the left, increasing the phase shift at any time following the reset, whilst increasing $t_u$ will shift the waveform to the right decreasing the phase shift.

The change of the phase of the signal at the load, $d\phi$ is:
\begin{equation}\label{eq:dPhi=-Omegadtau}
    d\phi=-\Omega d \tau
\end{equation}

From eq.\ref{eq:dPhi=-Omegadtau} it can be shown that if, for example, $\Omega=1\,\textrm{MHz}$, the error in $\tau$ would have to be less than $~2.7\,\textrm{ns}$ in order to achieve an error in the phase of less than $1^\circ$.  It will also be necessary to ensure that the maximum phase shift of the FE-FRT leaves enough overhead to allow for the largest phase correction required to be applied in a single reset.  If this is not the case a loss in conversion efficiency can occur via clipping of the phase waveform shown in fig.~\ref{fig:sawtooth}, as power from the source will not be frequency shifted for the duration of the clipping.

Feedback through switch timing allows correction to the phase to be made at a rep-rate of $~f_{rep}$.  For many use cases, $f_{rep}$ is expected to be fairly high, of the order of 1 MHz, this technique would therefore be more than fast enough for most applications.  A potentially significant advantage of this technique is that it does not require any additional power from the circuitry driving the FE-FRT(s).

Feedback by switch timing also allows another very interesting possibility: that of amplitude control of a resonant cavity.  Unlike the direct feedback method which only allows amplitude control in the magic-T circuit (sec.~\ref{sec:magic}), feedback by switch timing also allows amplitude control to be easily implemented in the circulator and series phase shifter systems (described in sec.~\ref{sec:Circulator} and \ref{sec:transmission} respectively).

Amplitude control via switch timing is achieved by setting $\Delta \Phi=\pi$ for a single reset, this is done by reducing $\tau$ by a factor of two. The phase of the signal at the cavity will then stay $\pi\,\textrm{rad}$ off the reference phase for a given number of periods, over which the usual $\Delta \Phi=2\pi$ will be maintained. During the out-of-phase state energy will be drained from the cavity at a rate faster than the natural decay time as the incoming signal will do work against the stored energy in the cavity.  Then another single period with $\Delta \Phi=\pi$ will revert the signal at the load to the normal drive phase. By alternating the width of time spent at the regular phase as compared to the 180 degrees off phase, the stored energy in a cavity with a sufficiently large decay time ($\gg\tau$) can thus be controlled.
 
\section{Magic-T based System}\label{sec:magic}

\begin{figure}[tb]
    \centering
\includegraphics[width=0.9\columnwidth]{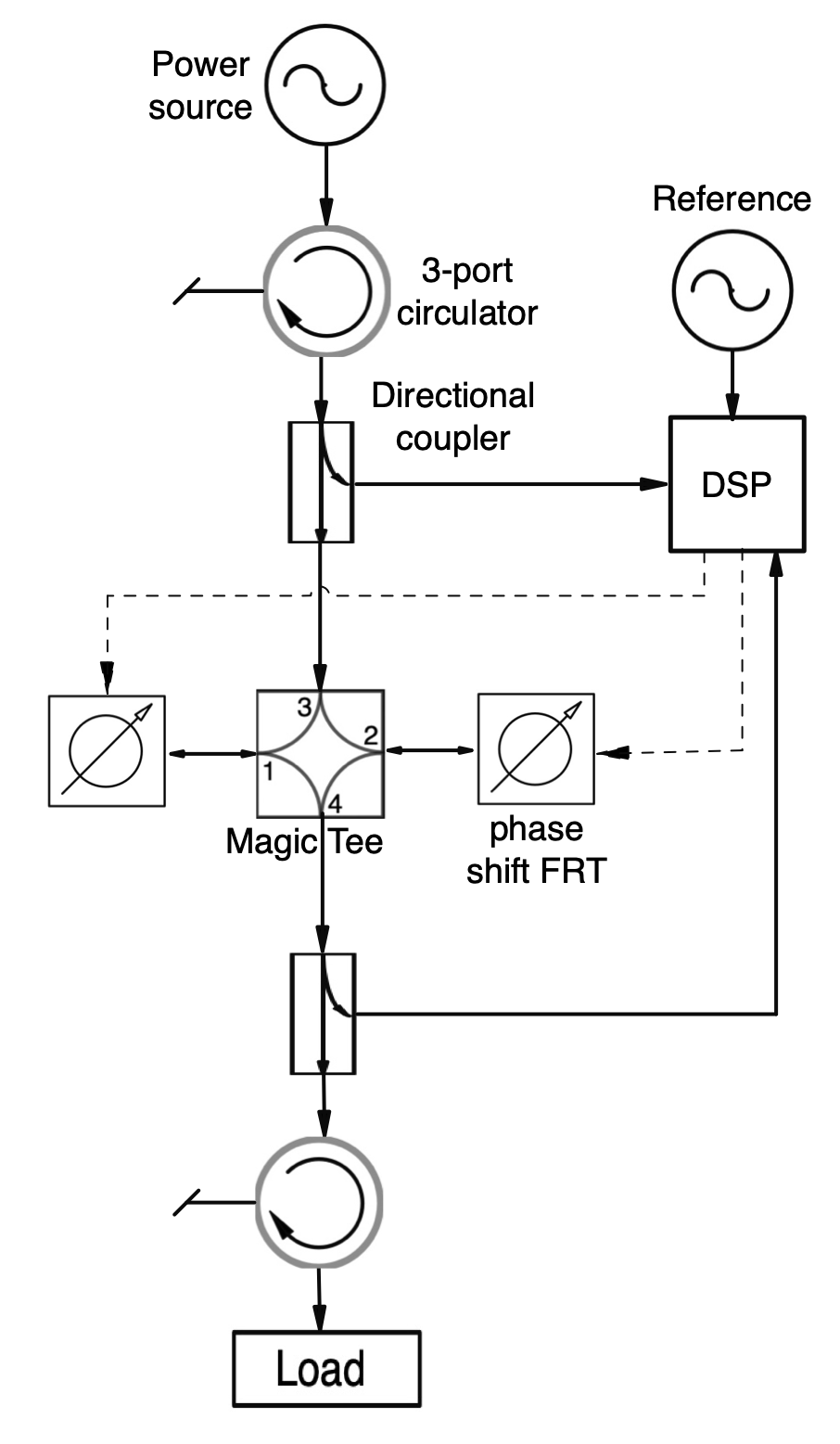}   
    \caption{Schematic diagram of the frequency modulator. The dashed lines represent low frequency control signals. The phase shifters are ferroelectric fast reactive tuners.}
     \label{magic}
\end{figure}

The schematic circuit for the previously mentioned magic-T based system is shown in fig.~\ref{magic}. The microwave source, protected by a circulator and sampled through a directional coupler is connected to the sum port (port-3) of a magic-T. A load is connected at the difference port (port-4); power reflected by the load is isolated by a second circulator. Two FE-FRT based phase shifters are connected to each of the collinear arms (ports 1 and 2).  

The circuit in fig.~\ref{magic} is similar to phase and amplitude control schemes proposed by: D. Valuch\cite{Valuch2004}, using ferrite based phase shifters; and Yakovlev, Kazakov and Hirshfield\cite{yakovlev20061}, using ferro-electric based phase shifters.  These references describe in detail how phase control can be achieved by varying the phase of the phase shifters in the same direction and amplitude control can be achieved by attenuating the output to the load by changing the phase of each phase shifter in opposite directions.  

The frequency modulation scheme proposed here can also be used for phase and amplitude control, as described in \cite{Valuch2004} and \cite{yakovlev20061} by superimposing the phase and amplitude control signal on top of the frequency modulation signal.

A full description of the magic-T is given by its 4 by 4 scattering matrix (S-matrix) shown in equation~\ref{eq:9} \cite{montgomery1965principles}. The order of the elements is first the co-linear arms (arms 1 and 2), then the sum port-3 and finally the difference port-4.
\begin{equation}\label{eq:9}
S=\frac{1}{\sqrt{2}}\begin{pmatrix}
0&0&1&1\\
0&0&1&-1\\
1&1&0&0\\
1&-1&0&0\\
\end{pmatrix}
\end{equation}
Designating the forward/returned voltages at the nth port of the magic-T as $V_n^+$/$V_n^-$, the return voltage vector is given by:

\begin{equation}
\begin{pmatrix}
V_1^-\\
V_2^-\\
V_3^-\\
V_4^-\\
\end{pmatrix}
=\frac{1}{\sqrt{2}}
\begin{pmatrix}
0&0&1&1\\
0&0&1&-1\\
1&1&0&0\\
1&-1&0&0\\
\end{pmatrix}
\begin{pmatrix}
V_1^+\\
V_2^+\\
V_3^+\\
V_4^+\\
\end{pmatrix}
    \label{eq:10}
\end{equation}

The RF power to be conditioned is applied to port-3. The forward voltages at ports 1 and 2 are the reflected voltages following a reflection at the FE-FRTs and traversing lengths of waveguide connecting each FE-FRT to the magic-T. For i=1,2:

\begin{equation}
V_i^+=\Gamma_i V_i^-
    \label{eq:11}
\end{equation}

From eq.~\ref{eq:2}, $\Gamma_i$ can be written as:

\begin{equation}
\Gamma_i \approx e^{j\theta_i}
    \label{eq:12}
\end{equation}

Then the following phases at the terminals of the two collinear arms are set as:
\begin{equation}
    \begin{aligned}[c]
       \Gamma_1  = e^{-j \Omega t}\\
          \Gamma_2  = e^{-j (\Omega t}+\pi)
    \end{aligned}
    \label{eq:13}
\end{equation}

The extra phase shift $e^{j\pi}$ in $\Gamma_2$ can be achieved by choosing the length of waveguide connecting arm-2 of the magic-T to its FE-FRT to be $\lambda/4$ longer than the waveguide connecting arm 1 of the magic-T to its FE-FRT. Thus:
\begin{equation}
    \begin{aligned}[c]
      V_1^+  = V_1^-e^{-j \Omega t}\\
          V_2^+  = V_2^-e^{-j (\Omega t}+\pi)
    \end{aligned}
    \label{eq:14}
\end{equation}
The forward (input from magnetron) signal at port-3 is given as:
\begin{equation}
V_3^+ =  V_M e^{j\omega_0 t}e^{j\Omega t}
    \label{eq:15}
\end{equation}

Any reflected power from the magic-T port-3 or the load attached to port-4 is removed by the circulators. Therefore:
\begin{equation}
V_4^+ = 0
    \label{eq:16}
\end{equation}

Looking at the two top rows of the scattering matrix, $V_1^-$ and $V_2^-$ are determined from $V_3^+$ and $V_4^+$. 
\begin{equation}
V_1^- = V_2^-=\frac{V_M}{\sqrt{2}}e^{j\omega_0 t}e^{j\Omega t}
    \label{eq:17}
\end{equation}
Therefore:
\begin{equation}
V_1^+ = -V_2^+=\frac{V_M}{\sqrt{2}}e^{j\omega_0 t}
    \label{eq:18}
\end{equation}
Thus, the input vector for the scattering matrix of the magic-T is determined:

\begin{equation}
\begin{pmatrix}
V_1^+\\
V_2^+\\
V_3^+\\
V_4^+\\
\end{pmatrix}
=V_M
\begin{pmatrix}
\frac{1}{\sqrt{2}}e^{j \omega_0 t}\\
-\frac{1}{\sqrt{2}}e^{j \omega_0 t}\\
e^{j \omega_0 t}e^{j \Omega t}\\
0\\
\end{pmatrix}
    \label{eq:19}
\end{equation}

Now multiply the scattering matrix by the input vector. To obtain the signal on port-4, multiply the input vector and the 4’th row of the S-matrix, to get $V_4^-$:

\begin{equation}
\begin{pmatrix}
V_1^-\\
V_2^-\\
V_3^-\\
V_4^-\\
\end{pmatrix}
=\frac{V_M}{\sqrt{2}}
\begin{pmatrix}
0&0&1&1\\
0&0&1&-1\\
1&1&0&0\\
1&-1&0&0\\
\end{pmatrix}
\begin{pmatrix}
\frac{1}{\sqrt{2}}e^{j \omega_0 t}\\
-\frac{1}{\sqrt{2}}e^{j \omega_0 t}\\
e^{j \omega_0 t}e^{j \Omega t}\\
0\\
\end{pmatrix}
    \label{eq:20}
\end{equation}
The result (other than insertion losses) shows that the full power of the magnetron appears at port-4 at the reference frequency:
\begin{equation}
V_4^-=V_M e^{j \omega_0 t}
    \label{eq:21}
\end{equation}
Furthermore, there is no power sent back to the magnetron at port-3:

\begin{equation}
V_3^-=0
    \label{eq:22}
\end{equation}
The above result has established that an RF source can be shifted to a desired reference frequency using a magic-T and ferroelectric controlled reactances.

\section{Circulator Based System}\label{sec:Circulator}
\begin{figure}[tb]
    \centering
\includegraphics[width=0.9\columnwidth]{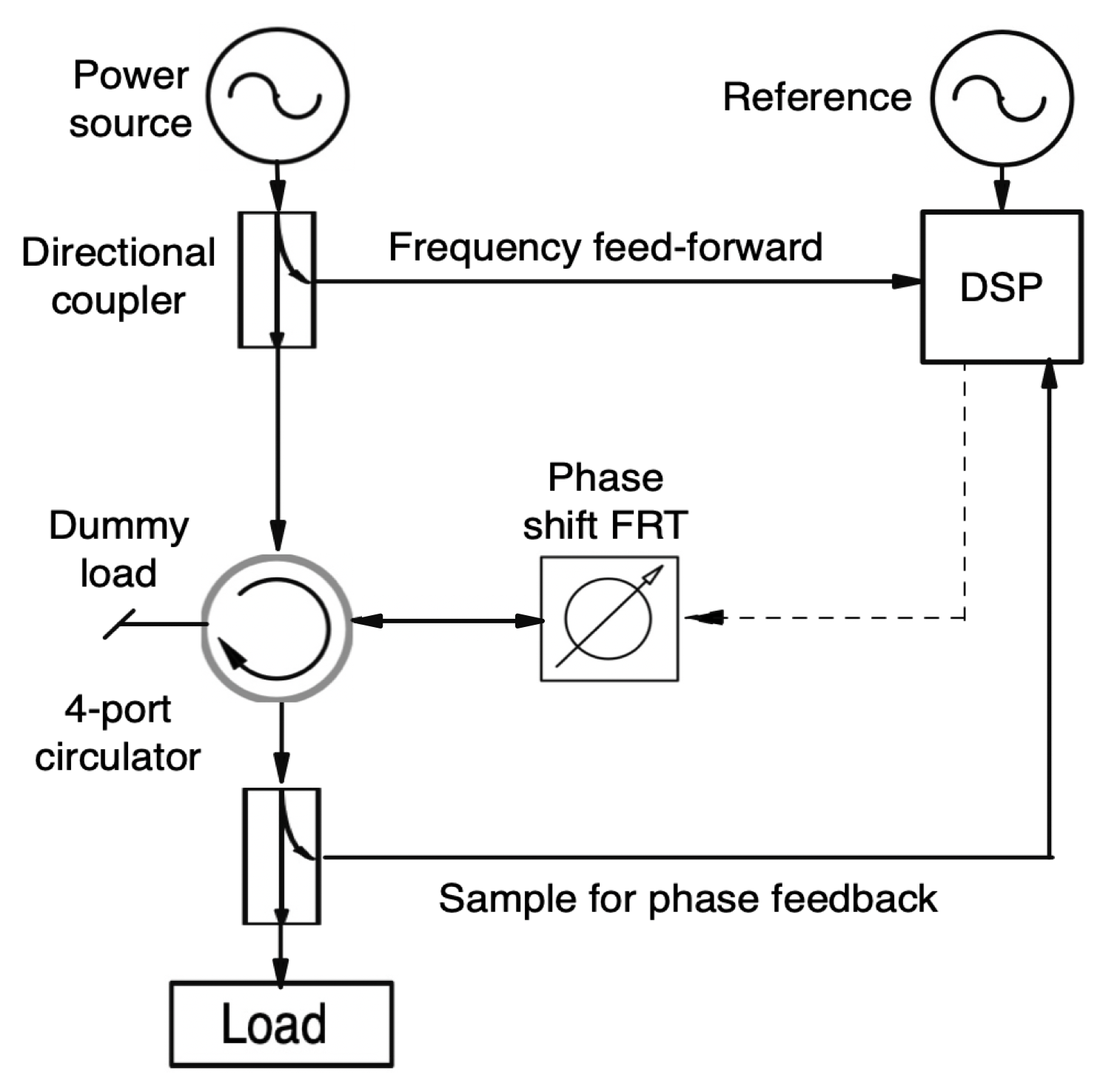}
    \caption{Schematic diagram of a circulator-based frequency modulator and fast phase feedback circuit. The dashed line represent low frequency control signals. The phase shifter is a single ferroelectric fast reactive tuner.}
    \label{circ}
\end{figure}

The second frequency modulation scheme is shown in Fig. \ref{circ}.
Power from the unstable source enters the 4-port circulator at the port-1, followed in the circulation direction by the FE-FRT phase shifter attached to port-2, the load at port-3 and a termination at port-4 of the circulator.
As before, the reflection coefficient at the FE-FRT phase shifter port can be written from eq.~\ref{eq:2} as:
\begin{equation}
\Gamma \approx e^{j\theta}
    \label{eq:23}
\end{equation}
where $\theta$ is the phase shift produced by the FE-FRT.  Also as before $\theta$ is set to be equal to $-\Omega t$.  Now, let the input voltage at port-1 be given as $V_1^+=V_Me^{j (\omega_0 +\Omega)t}$. Then as $V_2^-=V_1^+$ and $V_2^+=V_2^-\Gamma$ it follows that:
\begin{equation}
V_2^+= V_1^+\Gamma=V_Me^{j\omega_0t}
    \label{eq:24}
\end{equation}
Finally, as $V_3^-=V_2^+$ the signal for the load at port-3 is the required:
\begin{equation}
V_3^-=V_Me^{j\omega_0t}
    \label{eq:25}
\end{equation}

Thus it is established (ignoring the circulator insertion losses) that, provided $\theta=-\Omega t$, the desired frequency shift is added to the source signal which is then directed to the load.

As described in Section~\ref{sec:feedback}, phase and amplitude feedback can be implemented by controlling the length of the modulation cycle $\tau$.  Direct, fast and continuous phase feedback can also be applied by modulating the current driving the FE-FRT.

Compared to the magic-T circuit described in sec.~\ref{sec:magic}, the circulator circuit is significantly simpler, requiring half the number of FE-FRTs and fewer other components.  The total losses are likely to be lower, thus increasing efficiency.  On the other hand any FE-FRT will have to handle twice the power and direct amplitude feedback is not possible (only amplitude feedback by switch timing).

\section{Series Phase Shifter System}\label{sec:transmission}
\begin{figure}[tb]
    \centering
\includegraphics[width=0.9\columnwidth]{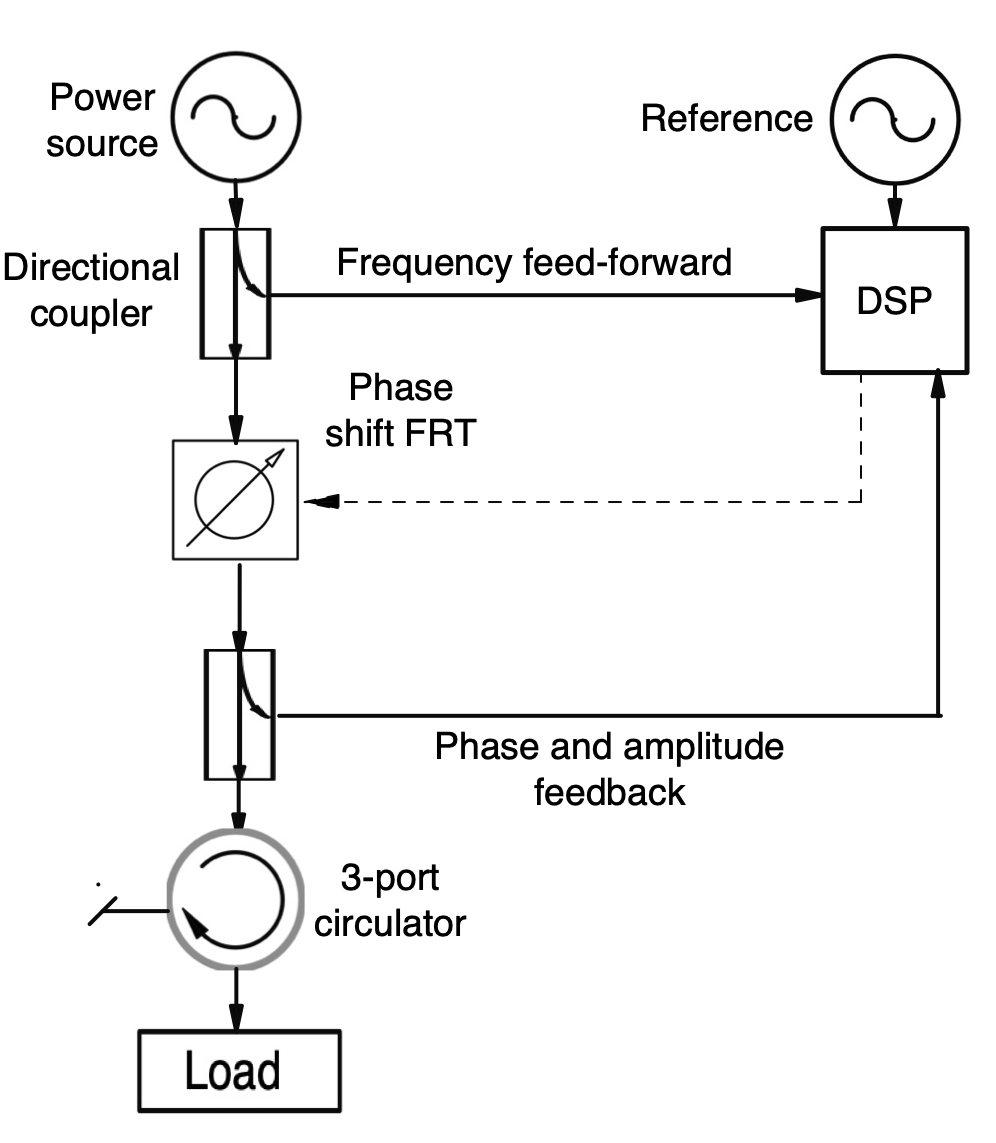}
    \caption{Schematic diagram of a frequency modulator and fast phase feedback circuit based on a transmission mode phase shifter. A circulator may be added to protect the source.}
    \label{trans_mode}
\end{figure}

The FE-FRT phase shifters described in Sections \ref{sec:magic} and \ref{sec:Circulator} are reflection based 1-port devices, inserting a refection coefficient in the manner of equation \ref{eq:2}. However, a transmission based 2-port phase shifter, placed in series between the source and load, could also be used as shown in Figure \ref{trans_mode}. 

In this configuration, the signal from the source is multiplied by the phasor $e^{j\theta}$, before being sent to the load.  Choosing again $\theta = -\Omega t$ then:
\begin{equation}
V_L=V_M e^{j \omega t}  e^{-j\Omega t}=V_M e^{j\omega_0 t}
    \label{eq:new26}
\end{equation}

This is the simplest of all the configurations and, as with the circulator based system, phase control can be achieved with both the direct and switch timing methods described in sec.~\ref{sec:feedback}, whereas amplitude control can only be achieved with the switch timing method.  As the source from the signal requires only one pass through a circulator before reaching the load, compared with two passes for the circulator based system, the losses should be lower resulting in an increased efficiency.  However, it remains to be shown that a low loss, high power, sufficiently fast, transmission style 2-port ferro-electric based phase shifter with a maximum phase shift equal to $\approx2\pi$ can be built.  Several 2-port ferro-electric phase shifters have been built and tested in the past\cite{Kazakov_2010}\cite{Wilson_2006}\cite{Flaviis_1997}\cite{Collier_1992} but the authors are not aware of any that have been built to date that would meet all of the requirements.  It is to be hoped that future research focused on developing a device which meets the requirements listed above will soon enable such a technology.

\section{The FE-FRT driver}\label{sec:driver}
The high-voltage driver is a critical component; it must be able to:
\begin{itemize}
\item Ramp up the voltage across the ferro-electric at the required rate
\item Quickly jump the voltage back to zero at the correct time
\item Correct for non-linearities in the phase as a function of applied voltage
\item Apply phase and amplitude corrections to the output signal.
\end{itemize}

\begin{figure}[tb]
    \centering
\includegraphics[width=0.95\columnwidth]{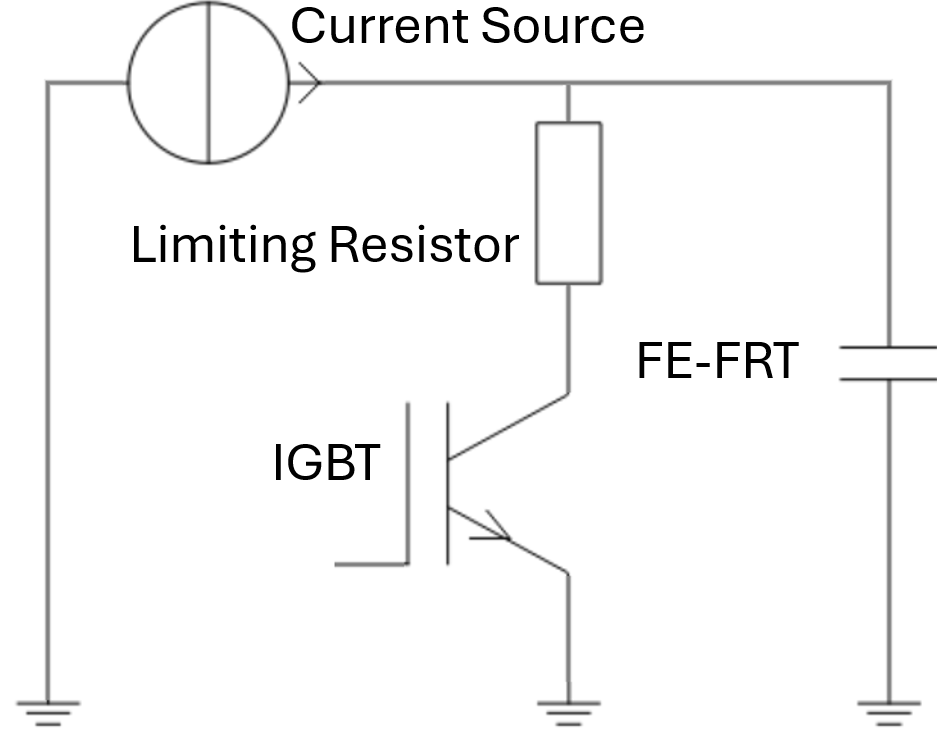}
    \caption{Conceptual circuit diagram to generate high voltage saw tooth waveform.}
\label{fig:HV_circuit}
\end{figure}

All of this can be achieved with the conceptual circuit shown in fig.~\ref{fig:HV_circuit}.  A voltage modulated current source provides an approximately constant current to charge the capacitance of the FE-FRT.  Modulating the current by a small amount but at a high frequency (higher than $\Omega$) would allow for direct feedback as described in sec.~\ref{sec:direct}, this would likely be required to correct for the non-linearities in the phase shift provided by the FE-FRT as a function of applied voltage.

The circuit also allows for rapid discharging of the FE-FRT through a current limiting resistor by closing a solid state switch.  The switch must be able to: achieve a rep-rate of $\Omega$; handle voltages in the $\textrm{kV}$ range; and re-open after closing in a time $<<\tau$.  An IGBT might be suitable for this purpose.  Phase and amplitude feedback using the switch timing method, as described in sec.~\ref{sec:switch timing}, can also be achieved if the triggering of the solid state switch is variable and can be controlled with sufficient accuracy.

The power required by FE-FRT driver can be significant and must be considered carefully as it, of course, contributes to the overall efficiency of the system.  The power required for operation of the driver can be estimated by summing the contributions from: the dissipation in any resistance, $R_w$, in the wire connecting the current source to the FE-FRT; and the power dissipated when discharging the FE-FRT through the current limiting resistor and solid state switch.  The power dissipated in the resistance $R_w$ will be:
\begin{equation}
P_{Rw} = I^2R_w
\end{equation}
where I is the constant current which will be given by:
\begin{equation}
I = \frac{CV}{t_u}
\end{equation}
with $C$ the capacitance of the FE-FRT and V the maximum voltage before the reset.  $P_{R_w}$ can then be written as:
\begin{equation}
P_{R_w} = \frac{C^2V^2R_w}{t_u^2}
\end{equation}
The energy dissipated during the discharge will be the energy stored on the capacitance of the FE-FRT before the reset given by:
\begin{equation}
U_d = \frac{CV}{2}
\end{equation}
the dissipated power is then:
\begin{equation}
P_d = \frac{CV}{2\tau}
\end{equation}
The total power required of the FE-FRT driver is thus:
\begin{equation}
\begin{aligned}\label{eq:Pdriver}
P_t &= P_{Rw}+P_d\\
    &\approx\frac{CV^2\Omega}{2}\left(1+2R_wC\Omega\right)
\end{aligned}
\end{equation}
provided $t_u\approx\tau$ as is anyway required for high RF conversion efficiency.  Note that any additional power required for direct feedback or variations in power required due to switch timing feedback have been ignored as they are difficult to quantify without considering a specific use case which is beyond the scope of this paper.

It is clear from eq.~\ref{eq:Pdriver} that more power is required for larger $\Omega$, $C$, $V$ and $R_{w}$.  Care should thus be taken to: design an FE-FRT to minimize $C$ and $V$; and design the driver to keep $R_w$ sufficiently low.  The values of $C$ and $V$ however are likely to face other constraints from, for example, the RF power handling required by the FE-FRT.  Other considerations that must be considered include the maximum safe power dissipation in the solid state switch and how best to achieve the modulation of the current source to enable direct feedback.  To fully assess the practicality for any specific use case a full and detailed design of both the FE-FRT and FE-FRT driver would be required.

\section{Conclusions}\label{sec:conclusion}
This manuscript describes a method to convert the frequency of a high-power microwave source, such as a magnetron, to a stable signal at precisely a given reference frequency. The frequency conversion can be applied using three methods: a magic-T based scheme, a 4-port circulator scheme and a direct transmission phase shifter scheme. All three schemes allow phase and amplitude feedback of the power sent to the load, allowing a highly stable power, locked to a reference (master oscillator). 
The system efficiency in converting the frequency is high, mostly associated with the small dissipation in the ferroelectric fast reactive phase shifter.
The system has a high frequency agility, allowing a rapid change in the frequency of the power sent to the load while keeping the power source undisturbed.

\begin{acknowledgments}
The authors gratefully acknowledge the support of colleagues on the related aspects of ferroelectric tuners and careful reading of the paper, in particular A. Macpherson (CERN) and E. Cruz Alaniz (no affiliation).
\end{acknowledgments}

\bibliography{bib}

\begin{thebibliography}{12}%
\makeatletter
\providecommand \@ifxundefined [1]{%
 \@ifx{#1\undefined}
}%
\providecommand \@ifnum [1]{%
 \ifnum #1\expandafter \@firstoftwo
 \else \expandafter \@secondoftwo
 \fi
}%
\providecommand \@ifx [1]{%
 \ifx #1\expandafter \@firstoftwo
 \else \expandafter \@secondoftwo
 \fi
}%
\providecommand \natexlab [1]{#1}%
\providecommand \enquote  [1]{``#1''}%
\providecommand \bibnamefont  [1]{#1}%
\providecommand \bibfnamefont [1]{#1}%
\providecommand \citenamefont [1]{#1}%
\providecommand \href@noop [0]{\@secondoftwo}%
\providecommand \href [0]{\begingroup \@sanitize@url \@href}%
\providecommand \@href[1]{\@@startlink{#1}\@@href}%
\providecommand \@@href[1]{\endgroup#1\@@endlink}%
\providecommand \@sanitize@url [0]{\catcode `\\12\catcode `\$12\catcode `\&12\catcode `\#12\catcode `\^12\catcode `\_12\catcode `\%12\relax}%
\providecommand \@@startlink[1]{}%
\providecommand \@@endlink[0]{}%
\providecommand \url  [0]{\begingroup\@sanitize@url \@url }%
\providecommand \@url [1]{\endgroup\@href {#1}{\urlprefix }}%
\providecommand \urlprefix  [0]{URL }%
\providecommand \Eprint [0]{\href }%
\providecommand \doibase [0]{https://doi.org/}%
\providecommand \selectlanguage [0]{\@gobble}%
\providecommand \bibinfo  [0]{\@secondoftwo}%
\providecommand \bibfield  [0]{\@secondoftwo}%
\providecommand \translation [1]{[#1]}%
\providecommand \BibitemOpen [0]{}%
\providecommand \bibitemStop [0]{}%
\providecommand \bibitemNoStop [0]{.\EOS\space}%
\providecommand \EOS [0]{\spacefactor3000\relax}%
\providecommand \BibitemShut  [1]{\csname bibitem#1\endcsname}%
\let\auto@bib@innerbib\@empty
\bibitem [{\citenamefont {Adler}(1946)}]{Injection}%
  \BibitemOpen
  \bibfield  {author} {\bibinfo {author} {\bibfnamefont {R.}~\bibnamefont {Adler}},\ }\bibfield  {title} {\bibinfo {title} {A study of locking phenomena in oscillators},\ }\href@noop {} {\bibfield  {journal} {\bibinfo  {journal} {Proceedings of the IRE}\ }\textbf {\bibinfo {volume} {34}},\ \bibinfo {pages} {351} (\bibinfo {year} {1946})}\BibitemShut {NoStop}%
\bibitem [{\citenamefont {Vyas}\ \emph {et~al.}(2016)\citenamefont {Vyas}, \citenamefont {Verma}, \citenamefont {Maurya},\ and\ \citenamefont {Singh}}]{vyas2016review}%
  \BibitemOpen
  \bibfield  {author} {\bibinfo {author} {\bibfnamefont {S.~K.}\ \bibnamefont {Vyas}}, \bibinfo {author} {\bibfnamefont {R.~K.}\ \bibnamefont {Verma}}, \bibinfo {author} {\bibfnamefont {S.}~\bibnamefont {Maurya}},\ and\ \bibinfo {author} {\bibfnamefont {V.}~\bibnamefont {Singh}},\ }\bibfield  {title} {\bibinfo {title} {Review of magnetron developments},\ }\href@noop {} {\bibfield  {journal} {\bibinfo  {journal} {Frequenz}\ }\textbf {\bibinfo {volume} {70}},\ \bibinfo {pages} {455} (\bibinfo {year} {2016})}\BibitemShut {NoStop}%
\bibitem [{\citenamefont {Ben-Zvi}\ \emph {et~al.}(2024)\citenamefont {Ben-Zvi}, \citenamefont {Burt}, \citenamefont {Castilla}, \citenamefont {Macpherson},\ and\ \citenamefont {Shipman}}]{ben2024conceptual}%
  \BibitemOpen
  \bibfield  {author} {\bibinfo {author} {\bibfnamefont {I.}~\bibnamefont {Ben-Zvi}}, \bibinfo {author} {\bibfnamefont {G.}~\bibnamefont {Burt}}, \bibinfo {author} {\bibfnamefont {A.}~\bibnamefont {Castilla}}, \bibinfo {author} {\bibfnamefont {A.}~\bibnamefont {Macpherson}},\ and\ \bibinfo {author} {\bibfnamefont {N.}~\bibnamefont {Shipman}},\ }\bibfield  {title} {\bibinfo {title} {Conceptual design of a high reactive-power ferroelectric fast reactive tuner},\ }\href@noop {} {\bibfield  {journal} {\bibinfo  {journal} {Physical Review Accelerators and Beams}\ }\textbf {\bibinfo {volume} {27}},\ \bibinfo {pages} {052001} (\bibinfo {year} {2024})}\BibitemShut {NoStop}%
\bibitem [{\citenamefont {Kazakevich}\ \emph {et~al.}(2018)\citenamefont {Kazakevich}, \citenamefont {Johnson}, \citenamefont {Lebedev}, \citenamefont {Yakovlev},\ and\ \citenamefont {Pavlov}}]{kazakevich2018resonant}%
  \BibitemOpen
  \bibfield  {author} {\bibinfo {author} {\bibfnamefont {G.}~\bibnamefont {Kazakevich}}, \bibinfo {author} {\bibfnamefont {R.}~\bibnamefont {Johnson}}, \bibinfo {author} {\bibfnamefont {V.}~\bibnamefont {Lebedev}}, \bibinfo {author} {\bibfnamefont {V.}~\bibnamefont {Yakovlev}},\ and\ \bibinfo {author} {\bibfnamefont {V.}~\bibnamefont {Pavlov}},\ }\bibfield  {title} {\bibinfo {title} {Resonant interaction of the electron beam with a synchronous wave in controlled magnetrons for high-current superconducting accelerators},\ }\href@noop {} {\bibfield  {journal} {\bibinfo  {journal} {Physical Review Accelerators and Beams}\ }\textbf {\bibinfo {volume} {21}},\ \bibinfo {pages} {062001} (\bibinfo {year} {2018})}\BibitemShut {NoStop}%
\bibitem [{\citenamefont {{Wang, H., et al.}}(2023)}]{WangSRF23}%
  \BibitemOpen
  \bibfield  {author} {\bibinfo {author} {\bibnamefont {{Wang, H., et al.}}},\ }\bibfield  {title} {\bibinfo {title} {{Demonstration of Magnetron as an Alternative RF Source for SRF Accelerators}},\ }in\ \href@noop {} {\emph {\bibinfo {booktitle} {Proc. SRF'23}}},\ \bibinfo {series and number} {International Conference on RF Superconductivity}\ (\bibinfo  {publisher} {JACoW Publishing, Geneva, Switzerland},\ \bibinfo {year} {2023})\ \bibinfo {note} {https://doi.org/10.18429/JACoW-SRF2023-WEPWB131}\BibitemShut {NoStop}%
\bibitem [{\citenamefont {Valuch}\ \emph {et~al.}(2004)\citenamefont {Valuch}, \citenamefont {Frischholz}, \citenamefont {T\"uckmantel},\ and\ \citenamefont {Weil}}]{Valuch2004}%
  \BibitemOpen
  \bibfield  {author} {\bibinfo {author} {\bibfnamefont {D.}~\bibnamefont {Valuch}}, \bibinfo {author} {\bibfnamefont {H.}~\bibnamefont {Frischholz}}, \bibinfo {author} {\bibfnamefont {J.}~\bibnamefont {T\"uckmantel}},\ and\ \bibinfo {author} {\bibfnamefont {C.}~\bibnamefont {Weil}},\ }\bibfield  {title} {\bibinfo {title} {First results with a fast phase and amplitude modulator for high power rf application},\ }\href@noop {} {\bibfield  {journal} {\bibinfo  {journal} {EPAC2004, Lucerne}\ ,\ \bibinfo {pages} {959}} (\bibinfo {year} {2004})}\BibitemShut {NoStop}%
\bibitem [{\citenamefont {Yakovlev}\ \emph {et~al.}(2006)\citenamefont {Yakovlev}, \citenamefont {Kazakov},\ and\ \citenamefont {Hirshfield}}]{yakovlev20061}%
  \BibitemOpen
  \bibfield  {author} {\bibinfo {author} {\bibfnamefont {V.~P.}\ \bibnamefont {Yakovlev}}, \bibinfo {author} {\bibfnamefont {S.~Y.}\ \bibnamefont {Kazakov}},\ and\ \bibinfo {author} {\bibfnamefont {J.}~\bibnamefont {Hirshfield}},\ }\bibfield  {title} {\bibinfo {title} {1.3 ghz electrically-controlled fast ferroelectric tuner},\ }\href@noop {} {\bibfield  {journal} {\bibinfo  {journal} {EPAC2006, Edinburgh}\ ,\ \bibinfo {pages} {487}} (\bibinfo {year} {2006})}\BibitemShut {NoStop}%
\bibitem [{\citenamefont {Montgomery}\ \emph {et~al.}(1965)\citenamefont {Montgomery}, \citenamefont {Dicke},\ and\ \citenamefont {Purcell}}]{montgomery1965principles}%
  \BibitemOpen
  \bibfield  {author} {\bibinfo {author} {\bibfnamefont {C.~G.}\ \bibnamefont {Montgomery}}, \bibinfo {author} {\bibfnamefont {R.~H.}\ \bibnamefont {Dicke}},\ and\ \bibinfo {author} {\bibfnamefont {E.~M.}\ \bibnamefont {Purcell}},\ }\href@noop {} {\emph {\bibinfo {title} {Principles of microwave circuits}}}\ (\bibinfo  {publisher} {Dover},\ \bibinfo {year} {1965})\ pp.\ \bibinfo {pages} {306--308}\BibitemShut {NoStop}%
\bibitem [{\citenamefont {Kazakov}\ \emph {et~al.}(2010)\citenamefont {Kazakov}, \citenamefont {Shchelkunov}, \citenamefont {Yakovlev}, \citenamefont {Kanareykin}, \citenamefont {Nenasheva},\ and\ \citenamefont {Hirshfield}}]{Kazakov_2010}%
  \BibitemOpen
  \bibfield  {author} {\bibinfo {author} {\bibfnamefont {S.~Y.}\ \bibnamefont {Kazakov}}, \bibinfo {author} {\bibfnamefont {S.~V.}\ \bibnamefont {Shchelkunov}}, \bibinfo {author} {\bibfnamefont {V.~P.}\ \bibnamefont {Yakovlev}}, \bibinfo {author} {\bibfnamefont {A.}~\bibnamefont {Kanareykin}}, \bibinfo {author} {\bibfnamefont {E.}~\bibnamefont {Nenasheva}},\ and\ \bibinfo {author} {\bibfnamefont {J.~L.}\ \bibnamefont {Hirshfield}},\ }\bibfield  {title} {\bibinfo {title} {Fast ferroelectric phase shifters for energy recovery linacs},\ }\href {https://doi.org/10.1103/PhysRevSTAB.13.113501} {\bibfield  {journal} {\bibinfo  {journal} {Phys. Rev. ST Accel. Beams}\ }\textbf {\bibinfo {volume} {13}},\ \bibinfo {pages} {113501} (\bibinfo {year} {2010})}\BibitemShut {NoStop}%
\bibitem [{\citenamefont {Wilson}\ \emph {et~al.}(2006)\citenamefont {Wilson}, \citenamefont {Fathy},\ and\ \citenamefont {Kang}}]{Wilson_2006}%
  \BibitemOpen
  \bibfield  {author} {\bibinfo {author} {\bibfnamefont {J.~L.}\ \bibnamefont {Wilson}}, \bibinfo {author} {\bibfnamefont {A.~E.}\ \bibnamefont {Fathy}},\ and\ \bibinfo {author} {\bibfnamefont {Y.~W.}\ \bibnamefont {Kang}},\ }\bibfield  {title} {\bibinfo {title} {{Investigation of Ferroelectrics for High-Power RF Phase Shifters in Accelerator Systems}},\ }in\ \href {https://jacow.org/l06/papers/THP030.pdf} {\emph {\bibinfo {booktitle} {Proc. LINAC'06}}},\ \bibinfo {series and number} {\bibinfo {series} {Linear Accelerator Conference}\ No.~\bibinfo {number} {23}}\ (\bibinfo  {publisher} {JACoW Publishing, Geneva, Switzerland},\ \bibinfo {year} {2006})\ pp.\ \bibinfo {pages} {637--639}\BibitemShut {NoStop}%
\bibitem [{\citenamefont {De~Flaviis}\ \emph {et~al.}(1997)\citenamefont {De~Flaviis}, \citenamefont {Alexopoulos},\ and\ \citenamefont {Stafsudd}}]{Flaviis_1997}%
  \BibitemOpen
  \bibfield  {author} {\bibinfo {author} {\bibfnamefont {F.}~\bibnamefont {De~Flaviis}}, \bibinfo {author} {\bibfnamefont {N.}~\bibnamefont {Alexopoulos}},\ and\ \bibinfo {author} {\bibfnamefont {O.}~\bibnamefont {Stafsudd}},\ }\bibfield  {title} {\bibinfo {title} {Planar microwave integrated phase-shifter design with high purity ferroelectric material},\ }\href {https://doi.org/10.1109/22.588610} {\bibfield  {journal} {\bibinfo  {journal} {IEEE Transactions on Microwave Theory and Techniques}\ }\textbf {\bibinfo {volume} {45}},\ \bibinfo {pages} {963} (\bibinfo {year} {1997})}\BibitemShut {NoStop}%
\bibitem [{\citenamefont {Collier}(1992)}]{Collier_1992}%
  \BibitemOpen
  \bibfield  {author} {\bibinfo {author} {\bibfnamefont {D.}~\bibnamefont {Collier}},\ }\bibfield  {title} {\bibinfo {title} {Ferroelectric phase shifters for phased array radar applications},\ }in\ \href {https://doi.org/10.1109/ISAF.1992.300662} {\emph {\bibinfo {booktitle} {ISAF '92: Proceedings of the Eighth IEEE International Symposium on Applications of Ferroelectrics}}}\ (\bibinfo {year} {1992})\ pp.\ \bibinfo {pages} {199--201}\BibitemShut {NoStop}%
\end{thebibliography}%

\end{document}